\documentclass[aps,prd,preprintnumbers,groupedaddress,nofootinbib,amssymb,eqsecnum,notitlepage]{revtex4}
\usepackage{graphicx}
\usepackage{bm}
\usepackage{amsmath}
\usepackage{color}
\usepackage{amsfonts}
\usepackage{here}
\usepackage{graphicx}
\usepackage{amsmath,amsthm,amssymb}
\usepackage{bm}
\allowdisplaybreaks[1]

\usepackage{amsfonts}
\usepackage{dcolumn}
\usepackage{hyperref}

\begin{document}
\newcommand{\newc}{\newcommand}

\newcommand{\ben}{\begin{eqnarray}}
\newcommand{\een}{\end{eqnarray}}
\newc{\be}{\begin{equation}}
\newc{\ee}{\end{equation}}
\newc{\ba}{\begin{eqnarray}}
\newc{\ea}{\end{eqnarray}}
\newc{\bea}{\begin{eqnarray*}}
\newc{\eea}{\end{eqnarray*}}
\newc{\D}{\partial}
\newc{\ie}{{\it i.e.} }
\newc{\eg}{{\it e.g.} }
\newc{\etc}{{\it etc.} }
\newc{\etal}{{\it et al.}}
\newcommand{\nn}{\nonumber}
\newc{\ra}{\rightarrow}
\newc{\lra}{\leftrightarrow}
\newc{\lsim}{\buildrel{<}\over{\sim}}
\newc{\gsim}{\buildrel{>}\over{\sim}}
\newc{\aP}{\alpha_{\rm P}}

\title{Hairy black hole solutions in $U(1)$ gauge-invariant 
scalar-vector-tensor theories}

\author{Lavinia Heisenberg$^{1}$
and Shinji Tsujikawa$^{2}$}

\affiliation{
$^1$Institute for Theoretical Studies, ETH Zurich, Clausiusstrasse 47, 8092 Zurich, Switzerland\\
$^2$Department of Physics, Faculty of Science, Tokyo University of Science, 1-3, Kagurazaka,
Shinjuku-ku, Tokyo 162-8601, Japan}

\date{\today}

\begin{abstract}

In $U(1)$ gauge-invariant scalar-vector-tensor theories with 
second-order equations of motion, we study the properties of 
black holes (BH) on a static and spherically symmetric 
background. In shift-symmetric theories invariant under 
the shift of scalar $\phi \to \phi+c$, we show the existence 
of new hairy BH solutions where a cubic-order scalar-vector interaction gives rise to a scalar hair manifesting 
itself around the event horizon. 
In the presence of a quartic-order interaction besides 
the cubic coupling, there are also regular BH solutions 
endowed with scalar and vector hairs.

\end{abstract}


\maketitle

\section{Introduction}
\label{introsec}

On the contrary to the geometrical interpretation of gravitational physics, the description in terms of field theory 
is unambiguous. It relies on the uniqueness of interactions of a massless spin-2 particle. The constructed Lagrangian 
inevitably leads to General Relativity (GR) with two propagating tensor degrees of freedom 
and with second-order equations of motion.

The extension from GR to modified gravity theories 
generally introduces new degrees of freedom besides 
two tensor polarizations \cite{review}. 
Under the assumptions of 
pseudo-Riemannian space-time, Lorentz symmetry, 
and locality, one can construct consistent tensor-tensor, vector-tensor and scalar-tensor theories with additional
tensor, vector or scalar fields in the gravity sector. 
Modified gravity theories based on an additional scalar field $\phi$
have been most extensively studied by reflecting their
simplicity. Fixing the ingredients of gravitational theory to be one spin-0 field besides two tensor polarizations, 
it is possible to construct most general scalar-tensor theories with second-order equations of motion, known as 
Horndeski theories \cite{Horndeski:1974wa,Horndeskire}.
The resulting action contains derivative and non-minimal couplings to gravity without inducing Ostragradski instabilities. 

Instead of a scalar field, one can introduce an additional 
spin-1 field into the gravity sector with a richer phenomenology due to the existence of intrinsic vector modes. Analogous to scalar-tensor Horndeski theories, it is possible 
to construct most general vector-tensor theories with second-order equations of motion. 
Upon imposing the $U(1)$ gauge invariance 
of the vector field $A_{\mu}$, Horndeski obtained 
a single nonminimal
coupling of the vector field to the double dual Riemann tensor \cite{Horndeski:1976gi} without vector derivative self-interactions. 
If one abandons the gauge invariance like the case of 
a massive vector filed, 
there are derivative and nonminimal couplings to gravity giving rise to generalized Proca theories \cite{Heisenberg:2014rta}. 
Even if the longitudinal mode of the vector field behaves as 
the Horndeski scalar field, there are two important purely 
intrinsic vector 
interactions with no scalar counterpart \cite{Heisenberg:2014rta,Jimenez:2016isa} 
(see also Refs.~\cite{VectorTensorTheories}). 
The relevance of these vector-tensor 
theories for cosmology \cite{VTcosmoearly,VTcosmology} and compact 
objects \cite{screening,VTastrophys1,VTastrophys2,VTastrophys3} 
has been already extensively studied in the literature.

One can unify these two important classes of Horndeski and generalized Proca theories into the framework of 
scalar-vector-tensor (SVT) theories. In Ref.~\cite{Heisenberg18}, the construction of 
SVT theories with second-order equations 
of motion was performed for both the $U(1)$ gauge-invariant and the non gauge-invariant cases.
The new degrees of freedom arising in SVT theories may be 
relevant to the physics of black holes, inflation, 
dark energy, dark matter and the generation of magnetic fields. 
In light of the detection of gravitational waves from 
BH and neutron star mergers \cite{GW1,GW2}, it is of interest to study whether or not 
some ``hairs'' associated with the new degrees of freedom  
arise on a strong gravitational background.

In this letter, we study BH solutions in $U(1)$ 
gauge-invariant SVT theories on a static and spherically symmetric background. 
We show that the existence of cubic-order 
scalar-vector interactions allows the possibility for 
realizing a nontrivial scalar-field configuration. 
In shift-symmetric theories where the Lagrangian is 
invariant under the shift 
$\phi \to \phi+c$, there exist new hairy BH solutions 
with scalar hair supported by the scalar-vector interaction.
We derive iterative solutions both in the vicinity 
of the horizon and at spatial infinity for the theories containing 
cubic and quartic interactions. We note that some BH solutions 
have been discussed in Ref.~\cite{Cha} for the 
quartic interaction, but we will show that the cubic 
interaction is crucially important for the existence 
of BHs with scalar hair. 
We will also numerically confirm the regularity 
of solutions outside the horizon exterior.

This letter is organized as follows.
In Sec.~\ref{modelsec}, we revisit $U(1)$ 
gauge-invariant SVT theories and present the background equations of motion on the static and spherically symmetric 
spacetime.
In Sec.~\ref{cubicsec}, we show the existence of regular BH solutions with scalar hair for a cubic-order coupling.
In Sec.~\ref{quarticsec}, we extend the analysis to the case in which quartic-order interactions are present besides the cubic coupling.
We conclude in Sec.~\ref{concludesec}.

\section{Gauge-invariant SVT theories and equations of motion}
\label{modelsec}

In Ref.~\cite{Heisenberg18}, the SVT theories were 
constructed for both $U(1)$ gauge-invariant and broken
gauge-invariant cases. In this letter, we will focus on 
the gauge-invariant case. 
The most general gauge-invariant action of SVT  theories with 
second-order equations of motion is expressed in the form 
\be
\mathcal{S}=\int d^4x \sqrt{-g}\left(\sum_{i=3}^5
\mathcal{L}^i_{\rm ST}+\sum_{i=2}^4\mathcal{L}^i_{\rm SVT}\right)\,,
\ee
where $g$ is a determinant of the metric tensor $g_{\mu \nu}$, and
$\mathcal{L}^3_{\rm ST}, \mathcal{L}^4_{\rm ST}, 
\mathcal{L}^5_{\rm ST}$ are the cubic, quartic, and quintic 
Lagrangians in pure scalar Horndeski theories with 
a scalar field $\phi$ \cite{Horndeski:1974wa,Horndeskire}.
The other Lagrangians 
$\mathcal{L}^{2}_{\rm SVT},\mathcal{L}^{3}_{\rm SVT}, \mathcal{L}^{4}_{\rm SVT}$ 
correspond to the genuine scalar-vector-tensor interactions, 
whose explicit forms are given, respectively, by 
\ba
\label{genLagrangianSVT}
\mathcal{L}^2_{\rm SVT}&=&f_2(\phi,X,F,\tilde{F},Y)\,, \\
\mathcal{L}^3_{\rm SVT}&=&\mathcal{M}_3^{\mu\nu}\nabla_\mu \nabla_\nu \phi\,, \\
\mathcal{L}^{4}_{\rm SVT}&=& \mathcal{M}_4^{\mu\nu\alpha\beta}\nabla_\mu\nabla_\alpha \phi\nabla_\nu\nabla_\beta\phi+f_4(\phi,X)L^{\mu\nu\alpha\beta}F_{\mu\nu}F_{\alpha\beta} \,,
\ea
where
\be
X=-\frac{1}{2} \nabla_{\mu} \phi  \nabla^{\mu} \phi\,,
\qquad
F=-\frac{1}{4} F_{\mu \nu} F^{\mu \nu}\,,
\qquad
\tilde{F}=-\frac{1}{4} F_{\mu \nu} \tilde{F}^{\mu \nu}\,,
\qquad 
Y=\nabla_{\mu} \phi \nabla_{\nu} \phi 
F^{\mu \alpha}{F^{\nu}}_{\alpha}\,,
\ee
with the gauge-invariant field strength 
$F_{\mu \nu}=\nabla_{\mu}A_{\nu}-\nabla_{\nu}A_{\mu}$ 
and the dual strength tensor $\tilde{F}^{\mu\nu}=\mathcal{E}^{\mu\nu\alpha\beta} F_{\alpha\beta}/2$. 
Here, $\mathcal{E}^{\mu\nu\alpha\beta}$ is the anti-symmetric Levi-Civita tensor satisfying the normalization  
$\mathcal{E}^{\mu\nu\alpha\beta}
\mathcal{E}_{\mu\nu\alpha\beta}=-4!$.
The rank-2 tensor $\mathcal{M}_3^{\mu\nu}$ in the 
Lagrangian $\mathcal{L}^3_{\rm SVT}$ is 
of the form 
\be
\mathcal{M}^{\mu\nu}_3=
\left[ f_3(\phi,X)g_{\rho\sigma}+\tilde{f}_3(\phi,X)
\nabla_\rho \phi\nabla_\sigma \phi \right] \tilde{F}^{\mu\rho}\tilde{F}^{\nu\sigma}\,,
\label{M3}
\ee
where $f_3$ and $\tilde{f}_3$ are functions of $\phi$ and $X$.
Similarly, the rank-4 tensor 
$\mathcal{M}_4^{\mu\nu\alpha\beta}$ 
is constrained to be 
\be
\mathcal{M}^{\mu\nu\alpha\beta}_4=\left[ 
\frac12f_{4,X}(\phi,X)+\tilde{f}_4(\phi) \right] 
\tilde{F}^{\mu\nu}\tilde{F}^{\alpha\beta}\,,
\ee
where $f_4$ is a function of $\phi$ and $X$ with 
the notation $f_{4,X} \equiv \partial f_4/\partial X$, while 
the function $\tilde{f}_4$ depends on $\phi$ alone.
The double dual Riemann tensor $L^{\mu \nu \alpha \beta}$ is constructed out of the Riemann tensor $R_{\rho\sigma\gamma\delta}$ as
\be
L^{\mu\nu\alpha\beta}=\frac{1}{4} \mathcal{E}^{\mu\nu\rho\sigma}
\mathcal{E}^{\alpha\beta\gamma\delta} R_{\rho\sigma\gamma\delta}\,.
\ee
By construction, these theories contain five propagating degrees of freedom 
(one scalar, two vectors, and two tensors). 
In the limit of a constant scalar field $\phi$ with 
$f_4=\text{constant}$, the Lagrangian 
${\cal L}_{\rm SVT}^4$ reduces to the gauge-invariant 
vector interaction $L^{\mu\nu\alpha\beta}F_{\mu\nu}F_{\alpha\beta}$ advocated by Horndeski 
in 1976 \cite{Horndeski:1976gi}.

In order to study the existence of new BH solutions on the static and  spherically symmetric background, 
we consider the following Ansatz for the line element 
\be
ds^2=-f(r) dt^{2} +h^{-1}(r) dr^{2}+ r^{2} d\Omega^2\,,\qquad
d\Omega^2=d\theta^{2}+\sin^{2}\theta\,d\varphi^{2}\,,
\label{metric_bg}
\ee
where $t$, $r$ and $\Omega$ stand for the time, 
radial, and angular coordinates, respectively,
and the functions $f$ and $h$ depend explicitly on the radial coordinate. 
We denote the horizon radius by $r_h$, which is defined such that $f(r_h)=h(r_h)=0$. 
Furthermore, we have $f(r)>0$ and $h(r)>0$ outside the event 
horizon ($r>r_h$).

For the background metric (\ref{metric_bg}), the scalar field is 
of the form $\phi=\phi(r)$.
The vector field has the temporal component $A_0$ and
the spatial part $A_i$. The spatial components can
be further decomposed into its transverse and longitudinal components 
as $A_i=A_i^{(T)}+\nabla_i \chi$, with $\nabla^{i}A_i^{(T)}=0$. 
Demanding the regularity of the vector field at $r=0$, 
the transverse mode $A_i^{(T)}$ 
has to vanish \cite{screening}. 
Thus, the vector-field profile compatible with 
the background metric (\ref{metric_bg}) is given by 
\be
A_{\mu}=\left( A_0(r), A_1(r), 0, 0 \right)\,,
\label{vector_ansatz}
\ee
with $A_1(r)=\chi'(r)$, where a prime denotes the derivative with respect to $r$. 
Because of the $U(1)$ gauge invariance, the longitudinal mode $A_1(r)$ does 
not contribute to the dynamics of the vector field for this background configuration.

Since the BH solutions in scalar-tensor and vector-tensor 
theories have been already extensively studied in the 
literature \cite{Hawking,VTastrophys1,VTastrophys2,Hui,stensorBH}, we will 
concentrate on the new scalar-vector-tensor interactions $\mathcal{L}^i_{\rm SVT}$ 
besides the Einstein-Hilbert Lagrangian $R$. 
Namely, we study the theories given by the action 
\be
\mathcal{S}=\int d^4x \sqrt{-g}\left(\frac{M_{\rm pl}^2}{2}R
+\sum_{i=2}^4\mathcal{L}^i_{\rm SVT}\right)\,,
\label{action}
\ee
where $M_{\rm pl}$ is the reduced Planck mass.
For the background field configuration explained above, 
the $\tilde{F}$ term in $\mathcal{L}^2_{\rm SVT}$ 
vanishes and the $Y$ term can be expressed 
in terms of $X$ and $F$, as $Y=4FX$.
Therefore, we will simply consider the function $f_2$ of 
the form $f_2(\phi,X,F)$. 
On the background (\ref{metric_bg}), the term proportional to 
$\tilde{f}_3(\phi,X)$ in Eq.~(\ref{M3}) also vanishes. 
Then, the action (\ref{action}) reduces to 
\be
\mathcal{S}=4\pi \int dt\,dr \left[ M_{\rm pl}^2 
\sqrt{\frac{f}{h}} \left( 1-h-rh' \right) 
+\sqrt{\frac{f}{h}} r^2f_2+\frac{2}{\sqrt{f}}
rh^{3/2}\phi' A_0'^2f_3
-\sqrt{\frac{h}{f}} A_0'^2 \{ 4(h-1)f_4
-h^2\phi'^2 (f_{4,X}+2\tilde{f}_4) \}
\right]\,.
\label{action2}
\ee

We recall that the functions $f_2, f_3, f_4, \tilde{f}_4$ have 
the dependence $f_2=f_2(\phi,X,F)$, $f_3=f_3(\phi,X)$, 
$f_4=f_4(\phi,X)$, and $\tilde{f}_4=\tilde{f}_4(\phi)$, where 
$X$ and $F$ are given, respectively, by 
\be
X=-\frac{h}{2}\phi'^2\,,\qquad 
F=\frac{h}{2f}A_0'^2\,.
\ee
Varying the action (\ref{action2}) with respect to 
$f,h,\phi,A_0$, respectively, the resulting equations 
of motion are
\ba
M_{\rm pl}^2 rfh' &=& M_{\rm pl}^2 f(1-h)
+r^2 \left( f f_2-hA_0'^2 f_{2,F} \right)
-2r h^2 \phi' A_0'^2f_3+hA_0'^2 \{ 
4(h-1)f_4-h^2\phi'^2 (f_{4,X}+2\tilde{f}_4) \}\,,\label{be1}\\
M_{\rm pl}^2 rh f' &=& M_{\rm pl}^2 f(1-h)
+r^2 \left( f f_2+fh \phi'^2 f_{2,X}-h A_0'^2 f_{2,F} 
\right)-2rh^2 \phi'A_0'^2 \left(3f_3-h \phi'^2 f_{3,X} \right) 
\nonumber \\
& &+hA_0'^2 \left[ 4(3h-1)f_4-h (9h-4)\phi'^2f_{4,X} 
+h^3 \phi'^4 f_{4,XX}-10h^2\phi'^2 \tilde{f}_4
\right]\,,\label{be2}\\
J_{\phi}' &=& {\cal P}_{\phi}\,,\label{be3}\\
J_{A}' &=& 0\,,\label{be4}
\ea
where we defined the following short-cut 
notations for convenience:
\ba
\hspace{-0.7cm}
J_{\phi} &=& -\sqrt{\frac{h}{f}} \left[  
r^2 f f_{2,X}\phi' -2h A_0'^2 (2h \tilde{f}_4+3hf_{4,X}-2f_{4,X})\phi' 
+2rh^2 A_0'^2f_{3,X}\phi'^2 
+h^3 A_0'^2 f_{4,XX}\phi'^3
-2rhA_0'^2 f_3 \right]\,,\label{Jphi}\\
\hspace{-0.7cm}
{\cal P}_{\phi} &=&
\frac{1}{\sqrt{fh}} \left[ r^2ff_{2,\phi}+hA_0'^2
\{ 4f_{4,\phi}+2h (r\phi'f_{3,\phi}-2f_{4,\phi})
+h^2 (f_{4,X\phi}+2\tilde{f}_{4,\phi})\phi'^2 
\} \right]\,,\\
\hspace{-0.7cm}
J_A &=& \sqrt{\frac{h}{f}} A_0' \left[ 
r^2 f_{2,F}+4rh\phi' f_3+8(1-h)f_4
+2h^2\phi'^2 (f_{4,X}+2\tilde{f}_4) \right]\,.
\label{JA}
\ea
{}From Eq.~(\ref{be4}), it follows that 
$J_A={\rm constant}$. 
This is attributed to the existence of a conserved 
charged current arising from the $U(1)$ gauge invariance.

In this letter, we will focus on shift-symmetric theories 
invariant under the shift 
\be
\phi \to  \phi+c\,,
\ee
where $c$ is a constant. 
Then, the functions $f_{2,3,4}$ and $\tilde{f}_4$ 
do not contain any $\phi$ dependence, such that 
\be
f_2=f_2 (X,F)\,,\qquad 
f_3=f_3 (X)\,,\qquad f_4=f_4(X)\,,\qquad 
\tilde{f}_4={\rm constant}\,.
\ee
Since ${\cal P}_{\phi}$ vanishes in this case,  
it follows that 
\be
J_{\phi}={\rm constant}\,,
\label{Jcon}
\ee
which means that the scalar equation of motion corresponds 
to the conservation of the current $J_{\phi}$. In this case, what we have is essentially the generalized Proca interactions
written in terms of scalar-vector-tensor theories with non-trivial
couplings between the scalar Stueckelberg field $\phi$ and the
gauge field $A_\mu$. The scalar Stueckelberg field enters only through derivatives.
In terms of the Stueckelberg field the cubic Lagrangian $\mathcal{L}^3_{\rm SVT}$ would correspond to the genuine vector interactions $g_5$ in generalized Proca theories, and the quartic Lagrangian $\mathcal{L}^4_{\rm SVT}$ to the genuine interactions in $\mathcal{L}_6$ of generalized Proca theories (see \cite{VTastrophys2} for the analysis of black hole solutions of generalized Proca interactions in the unitary gauge).

\section{Cubic interactions}
\label{cubicsec}

Let us first study BH solutions for the theories with 
$f_2 \neq 0$, $f_3 \neq 0$, $f_4=0$, and $\tilde{f}_4=0$. 
For concreteness, we consider the function $f_2$ given by 
the sum of $X$ and $F$, i.e.,  
\be
f_2(X,F)=X+F\,.
\ee
Then, the current (\ref{Jphi}) reduces to 
\be
J_{\phi}=-\sqrt{\frac{h}{f}} \left( r^2 f\phi'
+2rh^2A_0'^2f_{3,X}\phi'^2-2rhA_0'^2f_3 \right)\,.
\label{Jphi2}
\ee
We search for hairy BH solutions with finite values of  
$\phi'$ and $A_0'$ in the vicinity of the 
event horizon. We also consider the case in which the 
new scalar-vector interaction works as corrections to 
the Reissner-Nordstr\"{o}m (RN) metric of the form 
\be
f_{\rm RN}=h_{\rm RN}
=\left( 1-\frac{r_h}{r} \right) 
\left( 1-\mu \frac{r_h}{r} \right)\,,
\label{fRN}
\ee
where the constant $\mu$ is in the range $0<\mu<1$. 
Expanding the metric components $f$ and $h$ 
in the forms 
\be
f=\sum_{i=1}^{\infty}f_i \left( r-r_h \right)^i\,,\qquad
h=\sum_{i=1}^{\infty}h_i \left( r-r_h \right)^i\,,
\label{fh}
\ee
the leading-order RN solution (\ref{fRN}) corresponds to  
\be
f_1=h_1=\frac{1-\mu}{r_h}\,,
\ee
whereas the coefficients $f_i$ and $h_i$ with $i \geq 2$ 
are generally different from each other. 
Then, as $r \to r_h$, we have $\sqrt{h/f} \to 1$
in Eq.~(\ref{Jphi2}).

If we consider the function $f_3$ of the form 
$f_3=\beta_3 X^n$ with $n \geq 0$, the three terms 
in the parenthesis of Eq.~(\ref{Jphi2}) contain 
the positive powers of $f$ or $h$. 
This means that, as long as $\phi'$ and $A_0'$ are finite 
at $r=r_h$, the conserved current $J_{\phi}$ vanishes, 
so that 
\be
-\sqrt{\frac{h}{f}} \left( r^2 f\phi'
+2rh^2A_0'^2f_{3,X}\phi'^2-2rhA_0'^2f_3 \right)=0\,.
\label{Jphi3}
\ee
If the power $n$ is in the range $n \geq1$, the left hand side 
of Eq.~(\ref{Jphi3}) is factored out by $\phi'$. 
Then, the solution consistent with the boundary condition  
$\phi' \to 0$ at spatial infinity corresponds to $\phi'=0$ 
for arbitrary $r$, i.e, no scalar hair.
This situation is analogous to what happens in 
shift-symmetric Horndeski theories \cite{Hui}.

Instead, let us consider the cubic coupling 
\be
f_3(X)=\beta_3\,,
\label{f3cho}
\ee
where $\beta_3$ is a constant. 
In this case, the term $-2rhA_0'^2f_3$ in Eq.~(\ref{Jphi3}) 
does not contain $\phi'$. 
Then, from Eq.~(\ref{Jphi3}), we obtain 
the following solution 
\be
\phi'=\frac{2\beta_3h}{rf}A_0'^2\,.
\label{phiso}
\ee
This solution can be compatible with the boundary conditions 
$\phi' \to 0$ and $A_0' \to 0$ at spatial infinity. 
Substituting Eq.~(\ref{phiso}) and its $r$ derivative 
into Eqs.~(\ref{be1}), (\ref{be2}) and (\ref{be4}),
 it follows that 
\ba
M_{\rm pl}^2 rf h' &=& M_{\rm pl}^2f (1-h)
-\frac{h}{2f}A_0'^2 \left( fr^2+12 \beta_3^2 
h^2 A_0'^2 \right)\,,\label{f3be1}\\
M_{\rm pl}^2 rh f' &=& M_{\rm pl}^2f (1-h)
-\frac{h}{2f}A_0'^2 \left( fr^2+20 \beta_3^2 
h^2 A_0'^2 \right)\,,\\
A_0'' &=& -\frac{2 [M_{\rm pl}^2f^2 r^2
+4(1-h)\beta_3^2 fh M_{\rm pl}^2 A_0'^2-\beta_3^2 
h^2 r^2 A_0'^4]}{frM_{\rm pl}^2(fr^2+24\beta_3^2 
h^2 A_0'^2)}A_0'\,.\label{f3be3}
\ea
In the limit that $\beta_3 \to 0$, the solutions to 
Eqs.~(\ref{f3be1})-(\ref{f3be3}) are given by the 
RN metrics (\ref{fRN}) with the temporal 
vector component 
\be
A_0^{\rm RN}=P+\frac{Q}{r}\,,
\label{A0RN}
\ee
where $P$ and $Q$ are constants.

For $\beta_3 \neq 0$, we iteratively derive the solutions 
to Eqs.~(\ref{f3be1})-(\ref{f3be3}) both around the horizon 
and at spatial infinity.
Around $r=r_h$, we expand the two metric components 
of the form (\ref{fh}). The temporal vector component 
is also expanded as  
\be
A_0=a_0+\sum_{i=1}^{\infty}a_i \left( r-r_h 
\right)^i\,.
\label{A0ex}
\ee
Then, we obtain the following iterative solutions 
\ba
f &=&\left(1-\mu \right) \left( \frac{r}{r_h}-1 \right)
-\left[ 1-2\mu+12\tilde{\beta}_3^2 \mu^2  (1-\mu) \right]
\left( \frac{r}{r_h}-1 \right)^2+
{\cal O} \left( \frac{r}{r_h}-1 \right)^3\,,\label{fho}\\
h &=& \left(1-\mu \right) \left( \frac{r}{r_h}-1 \right)
-\left[ 1-2\mu-4\tilde{\beta}_3^2 \mu^2 (1-\mu)
 \right]
\left( \frac{r}{r_h}-1 \right)^2+
{\cal O} \left( \frac{r}{r_h}-1 \right)^3\,,\label{hho}\\
A_0 &=& a_0+\sqrt{2\mu}M_{\rm pl}
\left( \frac{r}{r_h}-1 \right)
-\sqrt{2\mu}M_{\rm pl} \left[ 
1+4 \tilde{\beta}_3^2\mu (2-\mu) \right]
\left( \frac{r}{r_h}-1 \right)^2
+{\cal O} \left( \frac{r}{r_h}-1 \right)^3\,,
\label{A0ho}
\ea
where $\tilde{\beta_3} \equiv \beta_3 M_{\rm pl}/r_h^2$, 
and we have chosen the branch $a_1>0$.
{}From Eq.~(\ref{phiso}), the field derivative is given by 
\be
\phi'=\frac{4 \tilde{\beta}_3\mu M_{\rm pl}}{r_h} 
\left[ 1- \left\{ 5+32 \tilde{\beta}_3^2  \mu
(1-\mu) \right\}
\left( \frac{r}{r_h}-1 \right)+
{\cal O} \left( \frac{r}{r_h}-1 \right)^2 \right]\,.
\label{phiho}
\ee
Thus, there exists a nontrivial scalar hair induced by 
the cubic scalar-vector coupling.
The coupling $\tilde{\beta}_3$ also gives rise to 
modifications to the metrics (\ref{fRN}) and the temporal 
vector component (\ref{A0RN}) of the RN solution.

To obtain the solutions at spatial infinity, we expand $f,h,A_0$ 
as the power series of $1/r$, as
\be
f=1+\sum_{i=1}^{\infty} \frac{\tilde{f}_i}{r^i}\,,\qquad
h=1+\sum_{i=1}^{\infty} \frac{\tilde{h}_i}{r^i}\,,\qquad
A_0=P+\sum_{i=1}^{\infty} \frac{\tilde{a}_i}{r^i}\,.
\label{larger}
\ee
Substituting these expressions into Eqs.~(\ref{f3be1})-(\ref{f3be3}), the resulting iterative solutions are 
\ba
f &=& 1-\frac{2M}{r}+\frac{Q^2}{2M_{\rm pl}^2r^2}
+\frac{3\beta_3^2Q^4}{14M_{\rm pl}^2r^8}
+{\cal O} \left( \frac{1}{r^9} \right)\,,\label{fla}\\
h &=& 1-\frac{2M}{r}+\frac{Q^2}{2M_{\rm pl}^2r^2}
-\frac{2\beta_3^2Q^4}{7M_{\rm pl}^2r^8}
+{\cal O} \left( \frac{1}{r^9} \right)\,,\\
A_0 &=& P+\frac{Q}{r}
-\frac{8\beta_3^2Q^3}{7r^7}
+\frac{2\beta_3^2MQ^3}{r^8}
+{\cal O} \left( \frac{1}{r^9} \right)\,,
\label{A0la}
\ea
where we have set $\tilde{f}_1=\tilde{h}_1=-2M$ and 
$\tilde{a}_1=Q$. On using Eq.~(\ref{phiso}), the scalar 
derivative behaves as
\be
\phi'=\frac{2\beta_3Q^2}{r^5}
+{\cal O} \left( \frac{1}{r^{11}} \right)\,,
\label{dphila}
\ee
which decreases rapidly toward the asymptotic value 0.
The effects of the coupling $\beta_3$ in $f$ and 
$h$ start to appear at the order of $1/r^8$, so the corrections 
to the RN metrics are suppressed to be small at spatial infinity. 
The correction to the RN value of $A_0$ arises at the 
order of $1/r^7$.

In Eqs.~(\ref{A0ho}) and (\ref{A0la}), both 
$a_0$ and $P$ are arbitrary constants. 
Indeed, they have no physical meanings
due to the $U(1)$ gauge symmetry.
At spatial infinity, there are two physical hairs $Q$ and $M$. 
Provided that the  BH solutions 
are regular throughout the horizon  
exterior, $Q$ and $M$ are related to the two parameters  
$\mu$ and $r_h$ in the vicinity of the horizon. 
Substituting the large-distance solutions (\ref{fla})-(\ref{A0la}) into the right hand side of Eq.~(\ref{JA}), 
it follows that the quantity $J_A$ is equivalent to 
the conserved $U(1)$ charge $-Q$. On the horizon, $J_A$ 
reduces to $\sqrt{2\mu}\,r_h M_{\rm pl}$, 
so the current conservation (\ref{be4}) gives the relation 
\be
\sqrt{2\mu}\,r_h M_{\rm pl}=-Q\,.
\label{Qrhre}
\ee
{}From Eq.~(\ref{dphila}), the scalar hair can be regarded as 
a secondary type sourced by the charge $Q$.

\begin{figure}[h]
\begin{center}
\includegraphics[height=3.4in,width=3.4in]{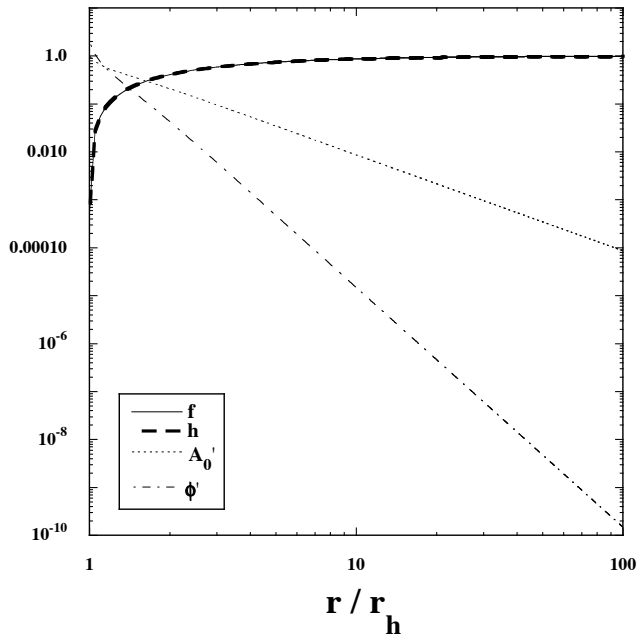}
\end{center}
\caption{\label{fig1}
Numerical solutions of $f,h,A_0',\phi'$ 
outside the event horizon for the cubic coupling 
$f_3=\beta_3$ with $\tilde{\beta}_3=\beta_3 M_{\rm pl}/r_h^2=1.0$ and $\mu=0.5$. 
Both $A_0'$ and $\phi'$ are normalized by 
$M_{\rm pl}/r_h$.
We choose the boundary conditions to be consistent 
with Eqs.~(\ref{fho})-(\ref{A0ho}) at $r=1.001r_h$.}
\end{figure}

To confirm the regularity of solutions outside the horizon,  
we numerically integrate 
Eqs.~(\ref{f3be1})-(\ref{f3be3}) with Eq.~(\ref{phiso}) 
by using the boundary conditions (\ref{fho})-(\ref{A0ho})
in the vicinity of the horizon. 
In Fig.~\ref{fig1}, we show the integrated solutions of 
$f,h,A_0',\phi'$ versus $r/r_h$ 
for $\tilde{\beta}_3=1.0$ and $\mu=0.5$. 
They are indeed regular throughout the horizon exterior.
The two metric components are close to 0 around $r=r_h$ 
and they asymptotically approach 1 for $r \gg r_h$. 
As estimated by Eqs.~(\ref{fho}) and (\ref{hho}), 
the cubic coupling $\tilde{\beta}_3$ induces the 
difference between $f$ and $h$ mostly around the horizon.
In the numerical simulation of Fig.~\ref{fig1}, 
the difference between $f$ and $h$ reaches a maximum  
close to the order $10^{-2}$ in the vicinity of the horizon and then $|f-h|$ 
decreases for increasing $r$.

{}From Eq.~(\ref{dphila}), the field derivative 
rapidly decreases as $\phi' \propto r^{-5}$ for $r \gg r_h$. 
This decreasing rate is larger than that of the derivative  
$A_0' \propto r^{-2}$.
Then, $\phi' \ll A_0'$ for $r \gg r_h$, but 
$\phi'$ cannot be neglected relative to $A_0'$ for 
$r$ closer to $r_h$.
For $\tilde{\beta}_3$ of the order of 1, the value 
$\phi'=4\tilde{\beta}_3 \mu M_{\rm pl}/r_h$ on the horizon 
becomes comparable to 
$A_0' \simeq \sqrt{2\mu}M_{\rm pl}/r_h$. 
Indeed, this property can be confirmed in Fig.~\ref{fig1}.
Thus, there exists the regular BH solution with a 
nontrivial scalar hair, whose effect mostly manifests in the 
nonlinear regime of gravity.

\section{Quartic interactions}
\label{quarticsec}

We proceed to the case of quartic interactions given by the 
Lagrangian ${\cal L}_{\rm SVT}^4$. 
First of all, we notice that each term in Eq.~(\ref{Jphi}) contains positive powers of $f$ and $h$. Hence, as long as  
$\phi'$ and $A_0'$ are finite on the horizon, the current 
$J_{\phi}$ is constrained to be 0. 
Moreover, except for the term $-2rh A_0'^2 f_3$, each term 
in $J_{\phi}$ is multiplied by the positive powers of $\phi'$.
This means that, for the theories with $f_3=0$, the 
solution to $J_{\phi}=0$ compatible with the 
boundary condition $\phi' \to 0$ at spatial infinity should 
correspond to the no scalar-hair solution $\phi'=0$ for 
arbitrary $r$.
Then, we need the cubic interaction for the realization 
of hairy BH solutions with nonvanishing $\phi'$.
In the following, we consider the model given by 
the functions 
\be
f_2(X,F)=X+F\,,\qquad 
f_3(X)=\beta_3\,,\qquad 
f_4(X)=\beta_4 X^n\,,\qquad 
\tilde{f}_4=0\,,
\ee
where $\beta_3, \beta_4$ and $n~(\geq 0)$ are constants.
For concreteness, we will study the two cases: 
(i) $n=0$ and (ii) $n=1$, separately. 

\subsection{Model with $f_4(X)=\beta_4$}

For the model with $n=0$, the current conservation 
$J_{\phi}=0$ leads to the relation same as 
Eq.~(\ref{phiso}) between $\phi'$ and $A_0'$.
On using this relation, Eqs.~(\ref{be1}), (\ref{be2}) and (\ref{be4}) reduce, respectively, to 
\ba
\hspace{-0.2cm}
M_{\rm pl}^2 rf h' &=& M_{\rm pl}^2f (1-h)
-\frac{h}{2f}A_0'^2 \left[ fr^2+12 \beta_3^2 
h^2 A_0'^2+8\beta_4 f (1-h) \right]\,,\label{f3be1d}\\
\hspace{-0.2cm}
M_{\rm pl}^2 rh f' &=& M_{\rm pl}^2f (1-h)
-\frac{h}{2f}A_0'^2 \left[ fr^2+20 \beta_3^2 
h^2 A_0'^2+8\beta_4 f (1-3h) \right]\,,\label{h3be1d}\\
\hspace{-0.2cm}
A_0'' &=& -\frac{2 [M_{\rm pl}^2f^2 r^2
+4(1-h)\beta_3^2 fh M_{\rm pl}^2 A_0'^2-\beta_3^2 
h^2 r^2 A_0'^4-4\beta_4 M_{\rm pl}^2f^2 (1-h) 
-8\beta_3^2 \beta_4 h^2(2h+1)A_0'^4]}{frM_{\rm pl}^2[fr^2+24\beta_3^2 
h^2 A_0'^2+8\beta_4 f (1-h)]}A_0'\,.\label{f3be3d}
\ea
Substituting Eqs.~(\ref{fh}) and (\ref{A0ex}) into Eqs.~(\ref{f3be1d})-(\ref{f3be3d}), 
the iterative solutions around $r=r_h$ up to the 
order of $(r-r_h)^2$ are given by 
\ba
\hspace{-0.7cm}
f &=&\left(1-\mu \right) \left( \frac{r}{r_h}-1 \right)
-\frac{1-2\mu+12 \tilde{\beta}_3^2 \mu^2 (1-\mu)
+4\tilde{\beta}_4 (24\tilde{\beta}_4 \mu^2-40\tilde{\beta}_4 
\mu+3\mu^2
+16\tilde{\beta}_4-9\mu+4)}{(1+8\tilde{\beta}_4)^2}
\left( \frac{r}{r_h}-1 \right)^2\,,\label{fho2}\\
\hspace{-0.7cm}
h &=& \left(1-\mu \right) \left( \frac{r}{r_h}-1 \right)
-\frac{1-2\mu-4 \tilde{\beta}_3^2 \mu^2 (1-\mu)
-4\tilde{\beta}_4 (8\tilde{\beta}_4 \mu^2
+8\tilde{\beta}_4 \mu+\mu^2
-16\tilde{\beta}_4+5\mu-4)}{(1+8\tilde{\beta}_4)^2}
\left( \frac{r}{r_h}-1 \right)^2\,,\label{hho2}\\
\hspace{-0.7cm}
A_0 &=& a_0+\sqrt{\frac{2\mu}{1+8\tilde{\beta}_4}}
M_{\rm pl}
\left( \frac{r}{r_h}-1 \right)
-\sqrt{\frac{2\mu}{(1+8\tilde{\beta}_4)^5}}M_{\rm pl} 
\left[ 1+4\tilde{\beta}_3^2 \mu (2-\mu) 
+4\tilde{\beta}_4-32\tilde{\beta}_4^2\right]
\left( \frac{r}{r_h}-1 \right)^2\,,
\label{A0ho2}
\ea
where $\tilde{\beta}_4 \equiv \beta_4/r_h^2$. 
{}From Eq.~(\ref{phiso}), the scalar derivative 
up to the order of $r-r_h$ reads
\be
\phi'=\frac{4 \tilde{\beta}_3\mu M_{\rm pl}}
{r_h(1+8\tilde{\beta}_4)} 
\left[ 1- \frac{5+32 \tilde{\beta}_3^2  \mu(1-\mu)
+16\tilde{\beta}_4 (2+\mu-4\tilde{\beta}_4
+8\tilde{\beta}_4 \mu)}{(1+8\tilde{\beta}_4)^2}
\left( \frac{r}{r_h}-1 \right) \right]\,.
\label{phiso1}
\ee

Substituting the large-distance expansions (\ref{larger}) into 
Eqs.~(\ref{f3be1d})-(\ref{f3be3d}) and setting 
$\tilde{f}_1=\tilde{h}_1=-2M$ and $\tilde{a}_1=Q$, 
the iterative solutions 
at spatial infinity yield 
\ba
f&=& 1-\frac{2M}{r}+\frac{Q^2}{2M_{\rm pl}^2 r^2}
-\frac{2\beta_4 Q^2}{M_{\rm pl}^2r^4}
+\frac{2\beta_4 MQ^2}{M_{\rm pl}^2r^5}
-\frac{3\beta_4 Q^4}{5M_{\rm pl}^4r^6}
+\frac{256\beta_4^2 MQ^2}{7M_{\rm pl}^2r^7}
\nonumber \\
& &
+\frac{3Q^2 (M_{\rm pl}^2Q^2 \beta_3^2 
-28\beta_4^2 Q^2-256\beta_4^2 M^2M_{\rm pl}^2)}{14M_{\rm pl}^4 r^8}\,,\label{fho2d}\\
h&=& 1-\frac{2M}{r}+\frac{Q^2}{2M_{\rm pl}^2 r^2}
-\frac{2\beta_4 M Q^2}{M_{\rm pl}^2 r^5}
+\frac{2\beta_4 Q^4}{5M_{\rm pl}^4 r^6}
-\frac{2Q^2 (\beta_3^2 Q^2-64\beta_4^2M^2)}
{7M_{\rm pl}^2 r^8}\,,\\
A_0 &=& P+\frac{Q}{r}-\frac{4\beta_4M Q}{r^4}
+\frac{3\beta_4 Q^3}{5M_{\rm pl}^2r^5}
-\frac{8Q (\beta_3^2 Q^2-32\beta_4^2M^2)}{7r^7}
+\frac{2MQ^3 (7\beta_3^2 M_{\rm pl}^2
-48 \beta_4^2)}{7M_{\rm pl}^2r^8}\,,
\label{A0so2}\\
\phi' &=& \frac{2\beta_3 Q^2}{r^5}
-\frac{64 \beta_3 \beta_4 MQ^2}{r^8}\,,
\label{phiso2}
\ea
up to the order of $1/r^8$.
{}From the current conservation (\ref{be4}), the $U(1)$ charge $Q$ 
at spatial infinity is related to the quantities $\mu$ and $r_h$ 
around the horizon as 
\be
\sqrt{2\mu (r_h^2+8\beta_4)}M_{\rm pl}=-Q\,.
\ee

In the limit that $\beta_3 \to 0$ the field derivatives 
(\ref{phiso1}) and (\ref{phiso2}) vanish, so there are no 
scalar hairs for the theories with $f_3=0$.
In this case, the large-distance solutions 
(\ref{fho2d})-(\ref{A0so2}) reduce to those 
with vector hair obtained by Horndeski in 1978 
for the pure $U(1)$-invariant interaction 
$\beta_4L^{\mu \nu \alpha \beta}
F_{\mu \nu}F_{\alpha \beta}$ \cite{HorndeskiBH}. 
We note that the iterative solutions (\ref{fho2})-(\ref{phiso1}) 
around the horizon are also consistent with those 
derived in Refs.~\cite{VTastrophys2}.
Unlike the solutions at spatial infinity arising from the 
cubic interaction, the quartic coupling 
$\beta_4$ leads to corrections to $f, h, A_0$ at the orders 
of $1/r^4, 1/r^5, 1/r^4$, respectively.
In the vicinity of the horizon, the coupling $\beta_4$ 
appears at the orders of $(r-r_h)^2$ in $f,h$ and 
of $r-r_h$ in $A_0$.

For $\beta_3 \neq 0$, the BH solutions 
derived above contain the nonvanishing scalar hair $\phi'$. 
Besides the quartic coupling $\beta_4$, the effects of 
$\beta_3$ on $f,h,A_0$ arise at the orders of $(r-r_h)^2$ around $r=r_h$, so that $f$ and $h$ are 
different from each other.
Compared to the case $\beta_4=0$, the derivative 
$\phi'$ on the horizon is modified by the factor 
$(1+8\tilde{\beta}_4)^{-1}$.
We have numerically integrated Eqs.~(\ref{f3be1d})-(\ref{f3be3d}) 
outside the horizon by using 
Eqs.~(\ref{fho2})-(\ref{phiso1}) as boundary conditions 
and found that the solutions smoothly connect to 
the large-distance solutions (\ref{fho2d})-(\ref{phiso2}) 
for $|\tilde{\beta}_4|<{\cal O}(0.1)$.
As in the case of Fig.~\ref{fig1}, the scalar hair 
manifests itself mostly in the vicinity of the horizon. 
Thus, there exist regular BH solutions endowed with both scalar and vector hairs. 
For $|\tilde{\beta}_4|>{\cal O}(0.1)$ the last terms in 
Eqs.~(\ref{f3be1d}) and (\ref{h3be1d}) can be larger 
than the first terms on their right hand sides, 
so that the regularity of BH solutions outside 
the horizon tends to be violated.

\subsection{Model with $f_4(X)=\beta_4X$}

Let us consider the quartic interaction $f_4(X)=\beta_4X$. 
{}From the current conservation $J_{\phi}=0$, 
it follows that 
\be
\phi'=\frac{2\beta_3rh A_0'^2}
{fr^2+2\beta_4h (2-3h)A_0'^2}\,,
\label{phiref}
\ee
which does not vanish for $\beta_3 \neq 0$.
The iterative solutions around the horizon  
expanded up to the order of $(r-r_h)^2$ 
are given by 
\ba
\hspace{-0.7cm}
f &=&\left(1-\mu \right) \left( \frac{r}{r_h}-1 \right)
-\frac{1-2\mu+12 \tilde{\beta}_3^2\mu^2 (1-\mu)
+8\bar{\beta}_4 \mu (1-2\mu)}{1+8\bar{\beta}_4 \mu}
\left( \frac{r}{r_h}-1 \right)^2\,,\label{fho3}\\
\hspace{-0.7cm}
h &=& \left(1-\mu \right) \left( \frac{r}{r_h}-1 \right)
-\frac{1-2\mu-4 \tilde{\beta}_3^2 \mu^2(1-\mu)
+8\bar{\beta}_4 \mu (1-2\mu)
}{1+8\bar{\beta}_4 \mu}
\left( \frac{r}{r_h}-1 \right)^2\,,\label{hho3}\\
\hspace{-0.7cm}
A_0 &=& a_0+\sqrt{2\mu}M_{\rm pl}
\left( \frac{r}{r_h}-1 \right)
-\sqrt{2\mu}M_{\rm pl}
\frac{
1+4\tilde{\beta}_3^2 \mu (2-\mu)
+16\bar{\beta}_4 \mu (1+2\tilde{\beta}_3^2 \mu
+4\bar{\beta}_4 \mu)}{(1+8\bar{\beta}_4 \mu)^2}
\left( \frac{r}{r_h}-1 \right)^2\,,
\label{A0ho3}
\ea
where $\bar{\beta}_4 \equiv \beta_4 M_{\rm pl}^2/r_h^4$.
{}From Eq.~(\ref{phiref}), the scalar derivative 
up to the order of $r-r_h$ yields
\be
\phi'= \frac{4\tilde{\beta}_3 \mu M_{\rm pl}}{r_h(1+8\bar{\beta}_4 \mu)} \left[ 1
-\frac{32\tilde{\beta}_3^2\mu (1-\mu)(1+4\bar{\beta}_4\mu)
+(1+8\bar{\beta}_4 \mu)^2 \{5+4\bar{\beta}_4 \mu
(3\mu-5) \}}{(1+8\bar{\beta}_4 \mu)^3} \left( \frac{r}{r_h}-1 \right) \right]\,.
\label{phirso}
\ee

At spatial infinity, we find that the solutions to $f,h,A_0$ 
are of the same 
forms as Eqs.~(\ref{fla})-(\ref{A0la}) up to the order of 
$1/r^8$. The scalar derivative $\phi'$ also 
has the same dependence 
as Eq.~(\ref{dphila}). Thus, unlike the constant $f_4$ model, 
the coupling $\beta_4$ does not appear in the large-distance 
expansions of $f,h,A_0,\phi'$ at the order lower than $1/r^8$.
Note that the current conservation (\ref{be4}) gives the relation 
same as Eq.~(\ref{Qrhre}).

Around the horizon, both the couplings $\beta_3$ and $\beta_4$ 
appear in Eqs.~(\ref{fho3})-(\ref{A0ho3}) at the order of 
$(r-r_h)^2$. From Eqs.~(\ref{phirso}), the scalar derivative $\phi'$ is comparable to 
$A_0' \simeq \sqrt{2\mu}M_{\rm pl}/r_h$ 
for $\tilde{\beta}_3/(1+8\bar{\beta}_4\mu)$ 
of the order of unity. 
Numerically we confirmed that the iterative solutions (\ref{fho3})-(\ref{phirso}) can connect to those at spatial infinity without discontinuities. 
Hence we have the regular BH solutions 
with the scalar hair manifesting itself 
in the vicinity of the horizon. 
We stress that this hairy solution arises by the presence 
of the cubic coupling $f_3=\beta_3$ besides the quartic interaction 
$f_4=\beta_4 X$.

\section{Conclusions}
\label{concludesec}

In this letter, we studied the static and spherically symmetric 
BH solutions in SVT theories with gauge-invariant derivative scalar-vector interactions and 
a non-minimal coupling to gravity. 
The longitudinal mode of the vector field does not 
propagate due to the $U(1)$ gauge invariance, so we are left 
with the equations of motion for a scalar field $\phi$ and 
a temporal vector component $A_0$ besides the 
gravitational equations of two metric components 
$f$ and $h$. In shift-symmetric theories where the functions 
$f_2,f_3, f_4,\tilde{f}_4$ in the action (\ref{action}) do 
not contain any $\phi$ dependence, there is the current 
conservation $J_{\phi}={\rm constant}$ associated 
with the scalar $\phi$.

Except for the term $-2rhA_0'^2 f_3$ in the square bracket 
of Eq.~(\ref{Jphi}), 
the current $J_{\phi}$ contains the products of 
positive powers of $\phi'$ and $f$ or $h$. 
The regularities of $\phi'$ and $A_0'$ on the horizon 
lead to $J_{\phi}=0$ in general.
Then, in the absence of the term $-2rhA_0'^2 f_3$, 
the solutions consistent with the boundary condition 
$\phi' \to 0$ at spatial infinity correspond to 
$\phi'=0$ for arbitrary radial distance $r$. 
Existence of the cubic interaction $f_3$ 
breaks this structure and allows the possibility for 
realizing hairy BH solutions with $\phi' \neq 0$.

In the presence of the cubic coupling $f_3=\beta_3$ 
besides the function $f_2=X+F$, the scalar derivative 
$\phi'$ is related to the temporal vector component 
in the form (\ref{phiso}). 
We derived the iterative solutions (\ref{fho})-(\ref{phiho}) expanded around the horizon and showed that the cubic coupling $\beta_3$ induces corrections to the RN solutions of 
$f,h,A_0$ at the order of $(r-r_h)^2$.
In the limit that $r \to r_h$, the scalar derivative 
approaches the nonvanishing finite value 
$\phi' \to 4\tilde{\beta}_3 \mu M_{\rm pl}/r_h$. 
At spatial infinity, the cubic coupling $\beta_3$ gives 
rise to corrections to $f,h,A_0'$ at the order of $1/r^8$. 
In this region, the scalar derivative quickly decreases 
as $\phi' \simeq 2\beta_3 Q^2/r^5$, 
but the scalar hair manifests itself around the horizon. 
Numerically, we confirmed the regularity of hairy BH 
solutions throughout the horizon exterior, see 
Fig.~\ref{fig1}. {}From the current conservation 
$J_A={\rm constant}$, the $U(1)$ charge $Q$ at spatial 
infinity is related to the quantities $\mu$ and $r_h$ 
around the horizon according to Eq.~(\ref{Qrhre}).

We also studied the cases in which the quartic interactions
$f_4=\beta_4X^n$ with $n=0$ or $n=1$ are present 
besides the cubic coupling $f_3=\beta_3$. 
In the limit that $\beta_3 \to 0$, the model with 
$f_4=\beta_4$ recovers BH solutions with 
vector hair discussed by Horndeski in 1978. 
Existence of the cubic coupling $\beta_3$ leads to 
a new hairy BH solution endowed with both scalar 
and vector hairs. For the quartic interaction $f_4=\beta_4X$, 
we also showed the presence of a hairy 
BH solution where the effects of couplings $\beta_3$ 
and $\beta_4$ on $f,h,A_0',\phi'$ manifests themselves
in the vicinity of the horizon.

In this letter, we showed the existence of hairy BH solutions 
induced by the scalar-vector interaction in $U(1)$ gauge-invariant SVT theories as a first step, but 
it is of interest to study what happens for 
the SVT theories with the broken gauge invariance. 
Moreover, the application of gauge-broken SVT theories 
to cosmology will be interesting in connection to the 
problems of inflation, dark energy, and dark matter. 
These topics are left for future works.

\section*{Acknowledgements}

We thank Masato Minamitsuji for useful comments 
and discussions.
LH thanks financial support from Dr.~Max R\"ossler, 
the Walter Haefner Foundation and the ETH Zurich
Foundation.  ST is supported by the Grant-in-Aid 
for Scientific Research 
Fund of the JSPS No.~16K05359 and 
MEXT KAKENHI Grant-in-Aid for 
Scientific Research on Innovative Areas ``Cosmic Acceleration'' (No.\,15H05890).


\end{document}